# Classification, Slippage, Failure and Discovery

Marc Böhlen[1]

[1] University at Buffalo, Department of Art, Emerging Practices in Computational Media
marcbohlen@protonmail.com

**Abstract.** This text argues for the potential of machine learning infused classification systems as vectors for a technically-engaged and constructive technology critique. The text describes this potential with several experiments in image data creation and neural network based classification. The text considers varying aspects of slippage in classification and considers the potential for discovery - as opposed to disaster - stemming from machine learning systems when they fail to perform as anticipated.

## 1 Introduction

New technologies bring with them the promise of new avenues for creative inquiry. Machine learning is no exception, yet it offers itself at the same time as a vehicle for testing the previous generalization. What kind of creative inquiry, if any, does machine learning in fact allow for, and which kinds of operations might this creative inquiry be applied to? Inversely, how might creative energies applied to machine learning modulate procedures and assumptions within machine learning itself?

The field of machine learning is wide and deep, and this short discussion concerns only one particular category of machine learning, namely supervised machine learning implemented with neural network architectures. Furthermore, the text investigates only machine learning supported image classification within visual culture.

## 2 Discovery

Broadly speaking, the primary contribution of machine learning is in algorithmic knowledge discovery: the automated detection of knowledge expressed as patterns and relationships in data. In a basic sense, supervised machine learning - detecting information patterns represented with a predefined collection of data - is a kind of search function that detects only what is already inherent in a dataset. Yet if the dataset is vast and distributed enough, seemingly trivial detection operations scale to something grander, as no human could practically perform this task.

Computer-based knowledge discovery is based on statistical methods (Piateski 1991) and can process vast amounts of data. Specifically, neural network-based machine learning was identified early on as a promising candidate for the development of knowledge discovery (Fu 1999). Over the past years, machine learning for knowledge discovery has been applied to domains ranging from materials science (Raccuglia 2016) to software design and vulnerability detection (Grieco 2016), drug and drug side effects research (Dimitri 2017) and the design of wining strategies for the game of GO (Silver 2018). In all of these examples, machine learning-supported methods of discovery generated useful and sometimes unexpected results that human-produced strategies were not able to create.

In the arts, discovery traditionally carries a different meaning. In particular, the Neoplatonist interpretation (Hendrix 2007) of the significance of the idea and imagination formed the groundwork for the unique position of artistic creativity that endures in several variations to the present. Through the power of *disegno inferno* (Zuccari 1607) the 'spark of fire', the artist perceives the artwork as a vision before it materializes into paint, stone (or code).

Computational creativity has, not without self-interest, attempted to expand and differentiate this traditional insular interpretation of creativity. Ventura, for example, places creative systems on a spectrum, beginning with the state of the merely generative (producing random events) through various states of filtration, inception and culminating in the act of creation at the end of the spectrum (Ventura 2016). In this framework, machinic discovery resides on the far end of the spectrum; distinct from what human-originating and unconstrained discovery abilities produce. However, the fact that the term discovery is now used across domains is at least indication of a contested territory within this generally accepted figuration of discovery. And as computing systems continuously expand the territory of what they can discover, they simultaneously suggest a re-evaluation of the uniqueness of human discovery dynamics as well as opportunities for collaborative human-computer discovery approaches.

# 3 Classification

Classification is omnipresent in everyday life. It is much older than the computational systems in which it is encoded today. In fact, to classify is human (Bowker 1999). More generally, classification is a form of controlled sense-making, a process of systematically composing predefined categories according to specific criteria. As opposed to generative systems, classification systems are by default forced to acknowledge their dependency on human agency in the very construction of categories. Classification is a supervised operation that requires information designers to formally represent the domain they attempt to capture in images, texts or sensor readings. These assemblages represent the world to the classification system.

The pedestrian world of machine learning-based classification offers no equivalent to the spectacular visual results that its younger cousin *Generative Adversarial Networks* can produce. Generative systems attract all the attention; they can create stunningly realistic portraits of people who never existed (Karras 2019), further eroding any vestiges of the notion of veracity in digital images. However, image classifications systems impact algorithmic and visual culture in their own unique ways. From the epistemology of defining what constitutes a category, to declaring what is 'found' when a classifier makes a decision (or a mistake), machine learning classification holds underexplored potential for inquiry. The following sections outline this argument through examples and considers some possible consequences for machine learning classification.

## 3.1 Image classification

Deep learning has become a potent method for learning from the world, and it has delivered state of the art results in voice recognition, sentiment analysis and image classification (LeCun 2015). Deep learning for image classification is a form of supervised deep learning in which a machine is exposed to examples (training data) of the objects it is tasked to recognize.

The most widely used network architectures for deep learning image classification are convolutional neural networks (CNNs). These networks apply a series of dimension reductions to ingest the multidimensional image data, creating a fully connected neural net. During training, a CNN is sequentially exposed to images in the training set and the corresponding category in a pattern of scores. The objective function measures the distance (error) between output scores and the desired pattern of scores. The machine modifies its multitude of internal parameters (weights) to reduce this error. The learning algorithm computes a gradient vector for each weight that indicates how the error would change if the weights were increased by a small value (LeCun p436). Adjustment of the weights then occurs in the opposite direction of that gradient, resulting in a complex adaptive negative feedback loop. This adjustment operation (backpropagation) is the 'magic sauce' of the learning process.  Eventually,  the average value of the objective function stops decreasing, and the network has 'learned' to represent its training data. At any given time, the weights represent the current knowledge of the network.

### 3.2 Interventionist image classification

The input data selection for network training provides substantial freedom of experimentation, as almost any kind of information can be applied to a neural network classifier.

But that opportunity comes with substantial commitment. It is very time-consuming and costly to design and create good image training sets. Not only must image categories be crisp enough for classifiers to robustly identify and distinguish, but the collection must also contain copious amounts of high-quality labeled data representing each of the categories. Not surprisingly, there exists a dearth of publicly available, professionally curated, high-quality datasets suitable for supervised machine learning. Moreover, many machine learning datasets suffer from a serious lack of diversity and a demonstrated bias in one form or another (Torralba 2011). And a data-labelling industry outsourced to operations using cheap labor presents its own obstacles to the quality control of image collections, as the case of the *CelebA* dataset has shown (Chandola 2017).

The dependencies of classifiers on input data are not only constraints, but opportunities. The configuration of a dataset is defined procedurally by the tasks which a classification system intends to perform, and it is always impacted by the biases and specific intentions of the design team putting the materials together in the first place. Hence, the most direct path into probing the logics and biases of neural networks is to confront them with data they were not specifically designed for. That is the approach Adam Harvey took in the project *Cluster Munition Detector Prototype* (Harvey 2018) which uses a Single Shot Detector (Liu 2016) approach to identify multiple objects in a single image. Instead of applying this technique to run-of-the-mill analytic tasks such as tracking human faces in a crowd or vehicles on a highway, Harvey applies his adaptation to the goal of finding munitions in the field, expanding the operational scope of this technique and facilitating at least in principle the dangerous work of mine-clearing crews.

As such, data-side interventions that leave the logics of networks unaltered yet bend the system into a different direction, constitute one new and viable approach to interventionist art. While this class of intervention is noteworthy, it can not disassociate itself from the logics of the network upon which it is constructed.

## 4 Accidents and dilemmas

The cultural critic Paul Virilio suggested that every invention creates its very own negativity, i.e. that the invention of the train co-invents the rail accident (Virilio 2007). Artificial intelligence systems are no exceptions to this observation.  In fact, artificially intelligent systems continue to create small and large accidents, a prominent example being a fatal traffic accident in which a self-driving vehicle failed to respond to a pedestrain pushing a bicycle across a road at night, killing the pedestrian (NTSB 2019).

In the context of this text, less spectacular examples of artificial intelligent accidents are perhaps more revealing of the main argument. A pertinent case is the report of an Asian man applying for a new passport, and being denied the document as the image analysis software determined that his eyes where not open, algorithmically re-creating an ugly ethnic stereotype (Cheng 2016). Such algorithm slippage demonstrates what can happen when laboratory-designed machine learning is confronted with real world data and cultural dynamics. Despite the fact that such incidents have become embarrassingly common, and despite the fact that both industry and academia openly acknowledge dysfunction in this class of machine learning, paths to ameliorating the problem remain stubbornly illusive.

While machine learning is subject to the *Collingridge dilemma* (Collingridge 1989), (Genus 2017) according to which technology control is difficult at early stages because not enough is understood about its consequences and costly later on, once the consequences are in fact apparent, it also occupies a particular space in the technology arena due to the dynamics of data sources upon which it relies. Supervised machine learning requires copious amounts of domain-specific and high-quality training data that is often only later, when a system is deployed in the field, found to be inadequate to the original task while the algorithm itself performs according to specifications.

## 5 Failure

Computer science provides an array of tools used to quantify the performance of neural network classifiers, from precision, recall and confusion scores to error rates (Tian 2020). These metrics help assess how well a classifier can detect a given category and how likely it is to confuse one category with another, for example.

In addition to these category-confusion approaches, researchers have developed a variety of algorithm-auditing procedures to assess which factors influence the performance of a given learning system. Algorithm audits study the functionality and the impact of algorithms. Computer scientists generally focus on the first aspect, functionality, while legal and humanities scholars focus on the second aspect, impact. Computer science has developed, for example, approaches by which one can estimate the degree to which a prediction would change had the model been fit on different training data (Schulam 2019). Humanities scholarship on the other hand has described how even algorithms that appear to work can be dangerous, because it may not be apparent when a breakdown occurs, and it is often unclear at what point they produce harm (Sandvig 2014). The lack of failure 'in obvious ways' in content personalization systems (Mittelstadt 2016) for example can lead to inadequate representation of content; this is a loss that the user remains ignorant of precisely because they never had the opportunity to be exposed to an alternate and possibly better solution.

## 6 Slippage

While quantitative measures are solid indicators of engineering performance, they are not tuned to uncover slippages in interpretation. Also, algorithm auditing studies tend to simplify complex data and decision landscapes in order to clarify cases for discrimination to a wide audience (Sandvig 2014). Moreover, the very concept of the audit comes with certain preconceptions. It already presupposes what kind of deficiencies it has to look for. Auditing implies that there is in fact something amiss that requires amelioration of one form or another.

The following sections take a different approach to supervised machine learning algorithm inquiry, focusing not on finding faults but on observing algorithmic behavior at the edges of their performance regime, highlighting in particular several small experiments via a mixed quantitative qualitative approach.

## 6.1 False similarities and unanticipated relationships

A variation of the analysis failure in the passport case mentioned above is the presence of 'false similarities'. False similarities are an aggregate effect of similarities across multiple features between test cases that make two categories appear more similar than they in fact are (Böhlen 2016). The false similarities approach was deployed, for example, to demonstrate how the output of the machine learning enabled expert system *Watson* - having detected in the authors of the *Communist Manifesto* very similar character traits as in the authors of an *IBM annual report* – suggested that the writers behind these radically different texts are much more similar than they could possibly be.

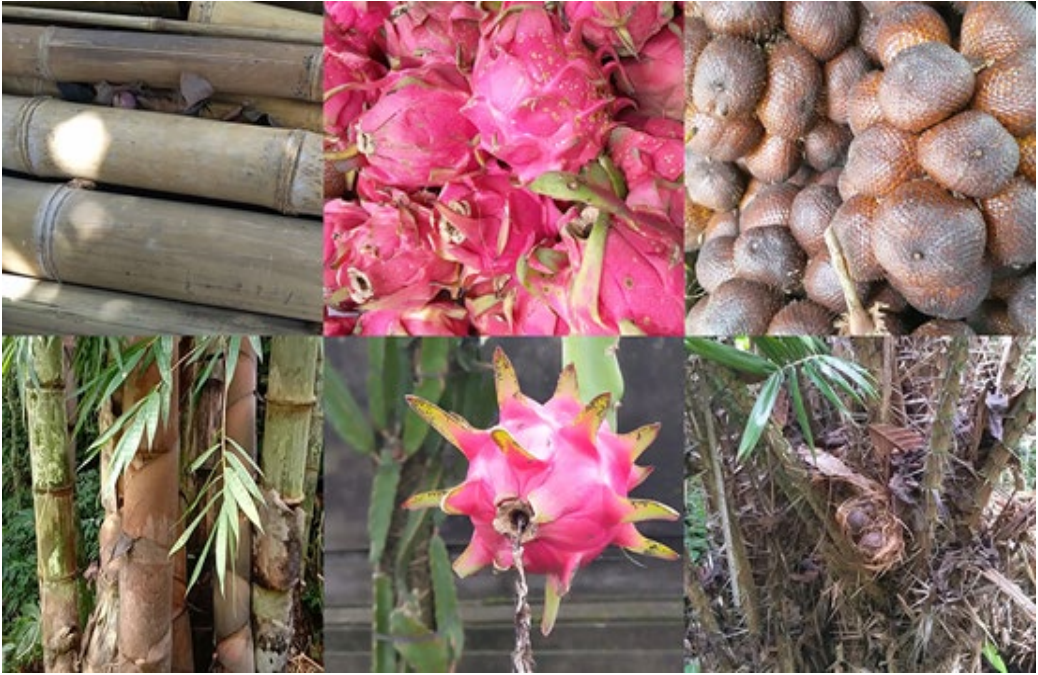

**Figure 1.** Images from the bali-26 collection (images courtesy of the author). Top row: bamboo at a construction site, dragon fruit and snake fruit at a market. Bottom row: bamboo, dragon fruit and snake fruit in the wild of Central Bali.

In addition to false similarities, one might consider more broadly 'unanticipated relationships' between classifier outputs. Unanticipated relationships in this context present themselves as classifier results that suggest commonalities across categories. An example of this type of classifier performance is described in the *Return to Bali* project[1] that aims to build a machine learning-compatible representation of ethnobotancially relevant plants from the island of Bali. Currently the collection comprises some 50'000 images of 26 distinct categories of flora collected in the wild from Central Bali. Of the half-dozen classifier architectures applied to the task, the ResNet152 network (He 2016) was able to identify the same plants in completely different contexts (see Fig. 1). Trained on images of bamboo, snake fruit and dragon fruit growing in forests, the network was able to detect bamboo on construction sites as well as snake fruit and dragon fruit at markets (Sujarwo 2020), showing that it was able to generalize across disparate contexts.

Detection across disparate contexts is a side-effect of neural networks' ability to generalize and find patterns across singular items in a way human beings often fail to. At the same time, the ability of networks to generalize can be fragile and fooled by noise that human observers disregard with ease

---

[1] http://www.realtechsupport.org/new_works/return2bali.html

(Serre 2019). Recently, researchers demonstrated the ability of neural network image classifiers to exhibit a form of generalization akin to Gestalt perception – the ability to perceive a whole only from parts (Kim 2019). The researchers found in several network topologies the ability to detect geometric forms such as triangles when only corner elements were presented to the system, an example of the Law of Closure (Wertheimer 1923). What makes this approach to network investigation significant in this context is not only the fact that the machine finds something we associate with human visual intelligence, but also that the experiment offers a new vector into probing the human side of Gestalt perception in the first place.

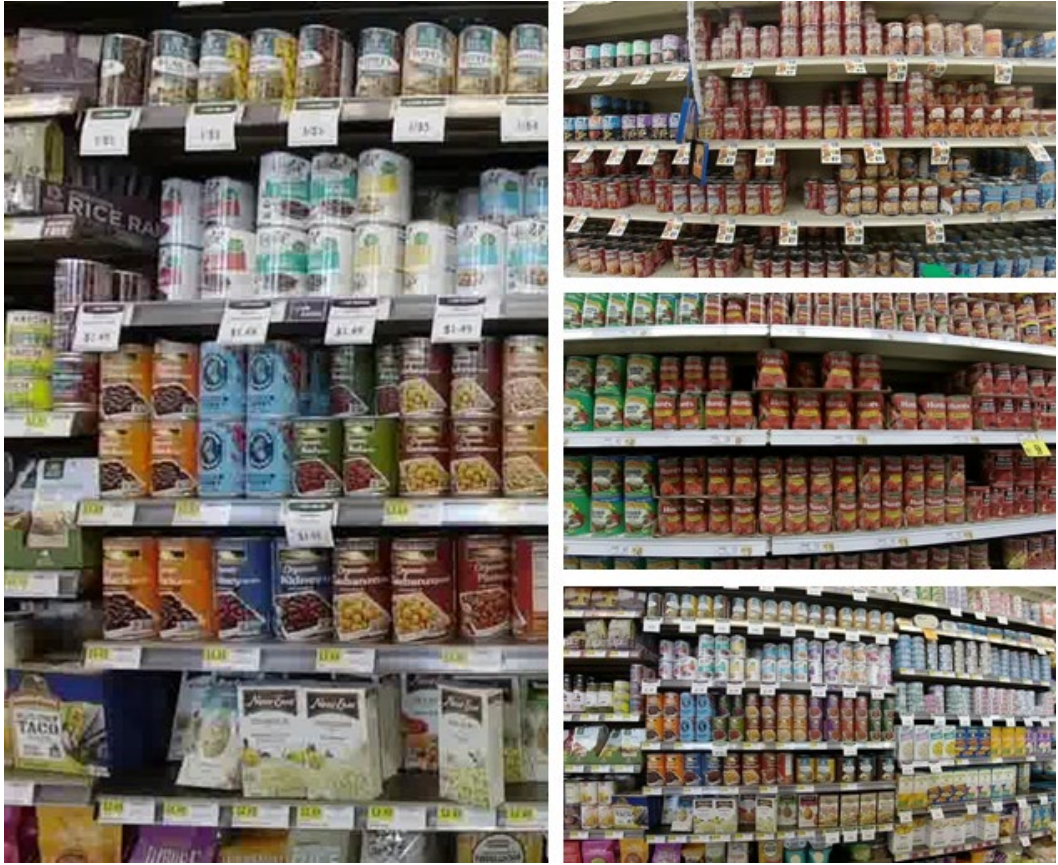

**Figure 2.** Identifying shopping goods with convolutional neural networks (images courtesy of the author). Left: personal-use sized canned goods at an upper scale grocery store. Right: arrangement of canned goods in three different grocery stores. Top: large sized canned goods. Center: large and mixed sized canned goods. Bottom: small sized canned goods.

# 7   Classification as discovery

At face value, discovery and classification may appear as mutually opposing operations. Discovery generally means the recognition of new relationships, and classification the assignment of pre-existing relationships. And yet it is possible that a trained classifier might 'know about' (encode) far more than the primary aim (classification categories) for which it was prepared. In other words, it has the potential for discovery because it is even 'smarter', i.e. it recognizes deeper patterns than it was trained for. This is a line of argumentation that artificial intelligence proponents might suggest. However, there is a different path along which discovery and classification meet, to wit, when classification fails.

```
/tops/cannedgoods_labeled/cans_13_16.jpg      predicted category: cannedgoods with 0.983491241932
/tops/cannedgoods_labeled/canned_21_6.jpg,    predicted category: beverages with 0.999880194664
/tops/cannedgoods_labeled/canned_10_17.jpg,   predicted category: cannedgoods with 0.965077579021
/tops/cannedgoods_labeled/goods_18_16.jpg,    predicted category: beverages with 0.988903701305
/tops/cannedgoods_labeled/goods_2_11.jpg,     predicted category: snacks with 0.616939246655
/tops/cannedgoods_labeled/canned_16_16.jpg,   predicted category: beverages with 0.547861039639
/tops/cannedgoods_labeled/Goods_14_8.jpg,     predicted category: cannedgoods with 0.977403461933
/tops/cannedgoods_labeled/goods_5_12.jpg,     predicted category: snacks with 0.429273635149
/tops/cannedgoods_labeled/canned_8_8.jpg,     predicted category: beverages with 0.873719930649
/tops/cannedgoods_labeled/goods_4_7.jpg,      predicted category: cannedgoods with 0.849367558956
```

**Figure 3**. Discovery in failure. Identifying shopping goods with convolutional neural networks (data courtesy of the author). The network trained on goods from one store struggles to detect the same category (of canned goods) in another store. Left side: ground truth (label), right side: classifier output showing failures (highlighted in red).

### 7.1 Discovery in failure - Lessons from the supermarket

As the observations on algorithm failure describe, the failure modes are not always obvious, and algorithms can fail even when they appear to succeed numerically. This last section describes the inverse phenomenon of 'failing while succeeding', namely 'succeeding while failing' and discusses the kinds of discovery that can occur in failure.

I recently collected several thousand images of grocery store items across five categories (beverages, canned goods, cereals, snacks and cleaning items) in preparation for the training of multiple neural network classifiers. These images were collected in three different grocery stores in Western New York. I then trained a generic low-dimensional convolutional neural network classifier on the images from one store and tested the trained model on images from the other two stores. Given the low-dimensionality of the network, the results were quite good, in the range of about 90% accuracy.

However, two categories faired much worse than the others, namely those of beverages and canned goods. While considering the reasons for the poor performance and checking the images themselves for clues, it became apparent what the 'failure' did indeed reveal, namely that canned goods and beverages varied considerably in product arrangement, size and shape across the three stores (see Fig. 2). In machine learning, the standard response to such a situation is to select a different classifier, one with a stronger discriminating capacity for the given task. Problem solved. In this case, an alternative response suggested itself. The apparent failure of the classifier could be scrutinized in a different way.

Canned goods in one store, catering to less-affluent members of the city were much more likely to be shelved in uniform larger sizes than in another store catering to health conscious and more affluent clientele who can afford the luxury of small packaged goods. What the failing classifier detected was an in-category difference in item packaging.

One store's canned goods are not another store's canned goods. Accounting for this observation, the system identified what the task had not asked for: merchandise difference across class boundaries: the social reality of cheap bulk produce landed outside of the features the simple classifier had assembled to identify canned goods.

Here, the system 'succeeds' in its failing mode as a mechanism that detects economic disparities behind the scenes, expressed in the differing visual dynamics of shelved goods. To be clear, this discovery requires some user support, as it were. It is co-produced by the classifier behavior and the observer who seeks to look beyond the immediate system response, namely the failure to classify 'properly' according to the pre-defined categories.

## 7.2 Something good can come from failing algorithms

Classifiers fail for many reasons, including poor concept and algorithm design, inadequate, incorrect, or sloppy data, and bias of every variety. More generally, however, they fail because they are tasked with an impossible challenge, to wit mapping complex realities onto simple outputs. As products of reductionist logics, classifiers are doomed to succumb to failure of one type or another.

The technical community goes to great lengths to address failure modalities, and the special status of failures in neural network classification is at least indirectly acknowledged through an explicit referral to the need for human expertise to handle 'hard cases' [Shi2020]. The discursive artificial intelligence community, on the other hand, identifies in failing algorithms proof of a fundamentally irrational project that algorithms are subscribed to, hopelessly entangled with the data-associated attributes of things and people (Amoore 2020).

Facing this binary landscape, I hope to locate an alternative, critical constructive position; a position that seeks a potential for discovery instead of the application of technological fixes, and that takes into account the fact that algorithms exist in ways that exceed their source code (Amoore 2020). So, I ask, how might one operationalize discovery in failure beyond the case I describe above? Unfortunately, the territory falls between the cracks of hard engineering work and eyes-wide-open reflexivity, suggesting that one must identify which professional class might even do this kind of work in the first place. Might the effort come from the side of regulatory and technology-savvy oversight practices such as *Public Interest Technology*[2]? Or more software-oriented algorithm auditing despite the tendency of that field to find what it – a priori - is looking for? Or might a different approach be in order, an approach that considers such underdetermined opportunities for discovery in fact as an act of creativity (as opposed to one of accounting)? It seems that the later would be necessary as discovery invariably requires some form of creative thinking and spontaneous acting under uncertain conditions. But which type of creativity could that be?

In the *Invention of Creativity*, the sociologist Reckwitz (Reckwitz 2017) offers a critical and harsh assessment of modern creativity, specifically of the personal gain-seeking variety. Reckwitz develops the concept of the 'creativity dispositif' as a constellation of coercive practices, modes of knowing and sensibilities circling around the production and institutionalization of relentless novel and 'exciting' innovation and 'being different'. In this arrangement, "the body, soul and practice become the self's own aesthetic object" (Reckwitz p222), largely excluding questions of shared concerns. Reckwitz offers at least one addition of interest to the endeavour I outline here, to wit *profane creativity*, a form of creativity that is not dependent on an audience and "locally situated". However, this profane creativity that "produces delights and discovery for the participants in the here and now" (Reckwitz p223) seems a bit too mild to address some of the tricky challenges machine learning systems tend to produce.

In the supermarket case described above, the dynamics between the failure of the machinery and the effort on the part of the viewer might be akin to the dynamics produced between a broken car and a driver when the dysfunctional vehicle transforms itself from annoyance into an opportunity to reflect on the vehicle in an utterly different way, as described in the *Tree of Knowledge* (Maturana 1987). A broken car with a dead battery then becomes a car with functioning open circuit, and an opportunity to reflect on open circuits in general, the invisible electronic control of complex consumer products, etc. The added value of this situated autopoetic process derives from the fact that it opens avenues for thoughts that might otherwise not occur.

In the case of the classification experiment described above, the system is not an isolated technical apparatus, and the resultant discovery carries with it considerable baggage, illuminating not only what we are not paying attention to, but also what we are reluctant to see in the first place. Because the issue (of unexpected indicators of economic inequality) only becomes apparent in the review of the algorithm's behavior and because the result speaks to an issue reflected in the source data, building a machine to 'solve the problem' seems unlikely to succeed.

---

[2] https://www.newamerica.org/pit/about/

Let's assume for argument's sake that the problem can be addressed (i.e. easily detected) with some very clever formalism, one that is able to make even a future artificial intelligence system accountable to communal values (Etzioni 2016). There is no guarantee that the system would also work properly on subsequent generations of yet more sophisticated machine learning systems. The algorithm-versus-critical reading encounter will repeat itself, requiring a new fix at each iteration in an infinite regress. Until superhuman artificial intelligence becomes a reality, human attention will have to intervene in the cycle in a yet to be defined artificial intelligence-meets-human intuition collaborative exchange.

In the meantime, an intermediate approach to developing 'revealing' (not merely transparent, accountable or interpretable) machine learning systems might be required. Clearly technical creative thinking is required. While there is no escaping the forces of the creativity dispositif when requiring some of its assets, it is possible to add new positions from outside of the creativity dispositive, as Reckwitz demonstrated. For example, can one include questions like "what kind of problems are we trying to solve?" (Stephensen 2020), and "can we recognize when new problems emerge from within the efforts to solve an existing one? "

And so we return to the question formulated at the onset regarding the scope and application-territory of creative inquiry in machine learning. Beyond the current infatuation with generative systems that portend to produce 'something new', the territory of discovery through failure outlined in this text deserves attention even though it does not belong to any particular formal type of inquiry. Maybe it is time to again to (or attempt to) update Agre's *Critical Technical Practice* (Agre 1997) for the age of ubiquitous, platform-centric, globally-distributed and data-heavy artificial intelligence to a less introspective, more collaborative approach distributed across multiple stakeholders; studying failures, developing algorithms while thinking deeply about what they enable, what kind of data they are dependent on, what they represent now and what they might come to represent in the future.

### 7.3 GitHub repository

The code originally created during the grocery store experiment and used to produce the image dataset in the *Return to Bali* project has been released as *Catch & Release*; an open source software package that facilitates audio-based image annotation for convolutional neural network label generation: *https://github.com/realtechsupport/c-plus-r*

# References

**Agre, Philip.** 1997. *Toward a Critical Technical Practice: Lessons Learned in Trying to Reform AI*. In: Geoffrey Bowker, Les Gasser, Leigh Star, and Bill Turner, eds, Bridging the Great Divide: Social Science, Technical Systems, and Cooperative Work, Erlbaum.
**Amoore, Louise.** 2020. *Cloud Ethics. Algorithms and the Attributes of Ourselves and Others*. Duke University Press.
**Böhlen, Marc.** 2016. *Watson Gets Personal. Notes on Ubiquitous Psychometrics*. xCoAx Conference Proceedings 2016, pp. 99-111.
**Bowker, Geoffrey, and Susan Leigh Star.** 1999. *Sorting Things Out*. MIT Press.
**Chandola, Varun, Marc Böhlen, and Amol Salunkhe.** 2017. *Server, server in the cloud. Who is the fairest in the crowd?* ArXiv:1711.08801v1.
**Cheng, Selina.** 2016. *An algorithm rejected an Asian man's passport photo for having "closed eyes"*. Quartz Magazine. December 7th. https://qz.com/857122/an-algorithm-rejected-an-asian-mans-passport-photo-for-having-closed-eyes/ Last accessed January 2021.

**Collingridge, David.** 1980. The Social Control of Technology. Pinter, London.
**Giovanna, Dimitri, and Lió Pietro.** 2017. DrugClust: *A machine learning approach for drugs side effects prediction*. Computational Biology and Chemistry, Volume 68, pp. 204-210.
**Etzioni, Amitai, and Oren Etzioni.** 2016. *AI Assisted Ethics.* Ethics and Information Technology, Vol. 18.2, pp. 149-156.
**Fu, LiMin.** 1999. *Knowledge discovery based on neural networks*. Commun. ACM 42, 11, pp. 47-50.
**Genus, Audley, and Andy Stirling.** 2017. Collingridge and the dilemma of control: towards responsible and accountable innovation. Research Policy, 47 (1), pp. 61-69.
**Grieco, Gustavo, Guillermo Grinblat, Lucas Uzal, Sanjay Rawat, Josselin Feist, and Laurent Mounier.** 2016. *Toward Large-Scale Vulnerability Discovery using Machine Learning.* CODASPY'16, New Orleans, USA.
**Harvey, Adam.** 2018. *Cluster Munition Detector Prototype*, https://vframe.io/research/cluster-munition-detector/ Last accessed January 2021.
**He, Kaiming, Xiangyu Zhang, Shaoqing Ren, and Jian Sun.** 2016. *Deep Residual Learning for Image Recognition*, 2016 IEEE Conference on Computer Vision and Pattern Recognition. Las Vegas, pp. 770-778.
**Hendrix, John.** 2007. *Humanism and Disegno: Neoplatonism at the Accademia di San Luca in Rome*. Bristol, Rhode Island: Roger Williams University School of Architecture, Art and Historic Preservation Faculty Publications.
**Karras, Tero, Samuli Laine, and Timo Aila.** 2019. *A Style-Based Generator Architecture for Generative Adversarial Networks*, 2019 IEEE/CVF Conference on Computer Vision and Pattern Recognition (CVPR), Long Beach, CA, USA, pp. 4396-4405.
**Kim, Been, Emily Reif, Martin Wattenberg, Samy Bengio, and Michael Moxer.** 2019. *Do Neural Networks Show Gestalt Phenomena? An Exploration of the Law of Closure*. ArXiv:1903.01069v3.
**Lecun, Yann, Yoshua Bengio, and Geoffrey Hinton.** 2015. *Deep learning*, Nature. 521(7553), pp. 436-444.
**Liu, Wei, Dragomir Anquelov, Dumitru Erhan, Christian Szegedy, Scott Reed, Cheng-Yang Fu, and Alexander Berg.** 2016. *SSD: Single Shot MultiBox Detector*. Lecture Notes in Computer Science: pp. 21-37.
**Maturana, Humberto, and Francisco Varela.** 1987. *The Tree of Knowledge*. Shambhala.
**Mittelstadt, Brent.** 2016. *Automation, Algorithms, and Politics. Auditing for Transparency in Content Personalization Systems*. International Journal of Communication, v. 10, p. 12, October.
**National Transportation Safety Board (NTSB).** 2019. *Preliminary Report HWY18MH010*. November 2019. https://dms.ntsb.gov/pubdms/search/hitlist.cfm?docketID=62978. Last accessed January 2021.
**Piateski, Gregory, and William Frawley.** 1991. *Knowledge Discovery in Databases*. MIT Press, Cambridge, MA, USA.
**Raccuglia, Paul, Katherine Elbert, Philip Adler, Casey Falk, Malia Wenny, Aurelio Mollo, Matthias Zeller, Sorelle Friedler, Joshua Schrier, and Alexander Norquist.** 2016. *Machine-learning-assisted materials discovery using failed experiments*. Nature 533, pp.73-76.
**Reckwitz, Andreas, and Steven Black.** 2017. *The Invention of Creativity: Modern Society and the Culture of the New*. Chicester: Polity Press.
**Sandvig, Christian, Kevin Hamilton, Karrie Karahalios, and Cedric Langbort.** 2014. *Auditing algorithms: Research methods for detecting discrimination on Internet platforms*. Paper presented at the 2014 International Communication Association Preconference on Data and Discrimination: Converting Critical Concerns into Productive Inquiry, Seattle, WA.
**Schulam, Peter, and Suchi Saria.** 2019. *Can You Trust This Prediction? Auditing Pointwise Reliability After Learning*. Proceedings of Machine Learning Research, in PMLR 89, pp.1022-1031.
**Serre, Thomas.** 2019. *Deep Learning: The Good, the Bad, and the Ugly*. Annu. Rev. Vis. Sci.5: pp. 399-426.
**Shi, Zheyuan Ryan, Clair Wang, and Fei Fang.** 2020. *Artificial Intelligence for Social Good: A Survey*. *ArXiv* abs/2001.01818.

**Silver, David, Thomas Hubert, Julian Schrittwieser, Ioannis Antonoglou, Matthew Lai, Arthur Guez, Marc Lanctot, Laurent Sifre, Dharshan Kumaran, Thore Graepel, Timothy Lillicrap, Karen Simonyan, and Demis Hassabis.** 2018. *A general reinforcement learning algorithm that masters chess, shogi, and Go through self-play*, Science, 07, pp. 1140-1144.

**Stephensen, Jan Løhmann.** 2020. *Post-creativity and AI: Reverse-engineering our Conceptual Landscapes of Creativity*. Proceedings of the Eleventh International Conference on Computational Creativity, pp. 326-332.

**Sujarwo, Wawan, and Marc Böhlen.** 2020. *Machine Learning in Ethnobotany*. IEEE International Conference on Systems, Man, and Cybernetics. Toronto, Canada, pp. 108-113.

**Tian, Yuchi, Ziyan Zhong, Vicente Ordonez, Gail Kaiser, and Baishakhi Ray.** 2020. Testing DNN image classifiers for confusion & bias errors. In Proceedings of the ACM/IEEE 42nd International Conference on Software Engineering (ICSE '20). Association for Computing Machinery, New York, NY, USA, pp. 1122-1134.

**Torralba, Antonio, and Alexei Efros.** 2011. *Unbiased look at dataset bias*, CVPR 2011, Providence, RI, pp. 1521-1528.

**Ventura, Dan.** 2016. *Mere Generation: Essential Barometer or Dated Concept*. Proceedings of the Seventh International Conference on Computational Creativity, pp. 22-29.

**Virilio, Paul.** 2007. *The Original Accident*. Cambridge: Polity.

**Wertheimer, Max.** 1923. *Laws of organization in perceptual forms*. First published as: Untersuchungen zur Lehre von der Gestalt II, in Psychologische Forschung, 4, pp. 301-350. Translation published in Ellis, W. 1938. A source book of Gestalt psychology, pp. 71-88, London: Routledge & Kegan Paul.

**Zuccari, Federico.** 1607. *L'Idea de' Pittori, Scultori e Architetti*, Torino. in Detlef Heikamp, ed., Scritti d'Arte di Federico Zuccaro (Firenze: Leo S. Olschki Editore, 1941).